\documentclass{emulateapj}

\usepackage{natbib}
\usepackage{rotate}
\usepackage{lscape}
\def\gtorder{\mathrel{\raise.3ex\hbox{$>$}\mkern-14mu
             \lower0.6ex\hbox{$\sim$}}}
\def\ltorder{\mathrel{\raise.3ex\hbox{$<$}\mkern-14mu
             \lower0.6ex\hbox{$\sim$}}}

\shorttitle{Discovery of 5000 AGNs behind the Magellanic Clouds}
\shortauthors{Koz{\l}owski \& Kochanek}

\begin{document}

\title{Discovery of 5000 Active Galactic Nuclei behind the Magellanic Clouds}

\author{Szymon~Koz{\l}owski\altaffilmark{1} and Christopher S. Kochanek\altaffilmark{1,2}
}

\altaffiltext{1}{Department of Astronomy, The Ohio State University, 140 West 18th Avenue, Columbus, OH 
43210; (simkoz, ckochanek)@astronomy.ohio-state.edu}
\altaffiltext{2}{The Center for Cosmology and Astroparticle Physics, The Ohio State University, 
191 West Woodruff Avenue, Columbus, OH 43210}

\begin{abstract}
We show that using mid-IR color selection to find Active Galactic Nuclei (AGNs) is as effective in dense stellar
fields such as the Magellanic Clouds as it is in extragalactic fields with low stellar
densities using comparisons between the {\it Spitzer} Deep, Wide-Field Survey data
for the NOAO Deep Wide Field Survey B\"ootes region and the SAGE Survey of the Large
Magellanic Cloud.  We use this to build high purity catalogs of $\sim 5000$ 
AGN candidates behind the Magellanic Clouds.  Once confirmed, these quasars will
expand the available astrometric reference sources for the Clouds and the numbers of
quasars with densely sampled, long-term ($>$decade) monitoring light curves by well over an 
order of magnitude and potentially identify sufficiently bright quasars for absorption
line studies of the interstellar medium of the Clouds.
\end{abstract}


%########################################################################
\section{Introduction}

At a first glance, searching for quasars behind the disk of the Galaxy or the Magellanic
Clouds seems neither useful nor promising.  While this presently holds for the problem of 
finding such quasars, they are useful tools once discovered.  In particular, they
serve as astrometric references for proper motion studies and the (ultraviolet) bright
quasars can be used to study absorption in the interstellar medium.  They may also be
unique tools for studying the variability of quasars.
After outlining the uses of such quasars and the results of existing searches for
them, we present a simple, robust procedure to
identify quasars in dense stellar fields with moderate extinction and generate a 
high purity catalog of quasar candidates in the Magellanic Clouds.

Our understanding of the dynamics of the Galaxy and the Large and Small Magellanic Clouds (LMC
and SMC) relies profoundly on our ability to measure the proper motions of stars, and a key use
of quasars in dense stellar fields is as astrometric references.  After early
efforts using ground based imaging in the Magellanic Clouds  
 (e.g., \citealt{1994AJ....107.1333,2000AJ....120..845A,2001AAS...199.5205D,2002AJ....123.1971P}) 
and the Galactic Bulge
  (e.g., \citealt{1992AJ....103..297S,2007MNRAS.378.1165R,2007AJ....134.1432V}),
most recent studies have used the higher astrometric precision of the {\it Hubble Space Telescope}
({\it HST}).  This has led to considerable recent progress on the proper motions of stars in the 
LMC (\citealt{2006ApJ...638..772K,2008AJ....135.1024P,2006AJ....131.1461P}), 
SMC (\citealt{2006ApJ...652.1213K})
and the Galactic bulge (\citealt{2002AJ....124.2054K,2006MNRAS.370..435K,2008ApJ...684.1110C}).
These studies are limited by the small numbers of available quasars, with less than 100 known for the LMC
(e.g., \citealt{2002AcA....52..241E,2003AJ....125....1G,2002ApJ...569..15D,2005A&A...442..495D}) 
and only a few dozen candidates behind the Galactic Bulge (e.g., \citealt{2005MNRAS.356..331S}).

A second use for these quasars is for absorption line studies of the interstellar medium
of the Galaxy and the Clouds (e.g., \citealt{2000ApJ...538L..73S,2000ApJS..129..563S}).  
For the Clouds this has largely been limited to studies of the
Magellanic Stream (e.g., \citealt{2000ApJ...538L..31S, 2005A&A...443..525S,2008ApJ...678..219L,2009arXiv0902.0208M}) 
rather than the Clouds themselves.  While there are no guarantees that sufficiently bright quasars exist behind
the Clouds, the advent of the Cosmic Origins Spectrograph for {\it HST} will allow the use of
significantly fainter quasars than earlier studies.

The third, and least obvious, use of these quasars is as the best existing data base for understanding
quasar variability.  Because of the microlensing studies of the Magellanic Clouds and the Galactic
bulge (EROS -- \citealt{1999A&A...344L..63A}; MACHO -- \citealt{2000ApJ...542..281A};  
MOA -- \citealt{2001MNRAS.327..868}; OGLE -- \citealt{2003AcA....53...291U}), most quasars 
identified in these regions will have well-sampled, decade long light
curves that can be used to study the optical variability of quasars.
The ensemble variability of quasars is well studied (e.g., \citealt{2003AJ....126.1217D, 2004ApJ...601..692V}), 
but less is known about the variability properties of 
individual quasars because it has been difficult to monitor them in large numbers.  While 
some newer projects will remedy this problem (e.g., QUEST, Ringstorf et al. 2009; Pan-STARRS, LSST),   
these data already 
exist for quasars in the microlensing regions and, to a lesser extent, for Stripe 82 of the Sloan Digital 
Sky Survey (SDSS, \citealt{2007MNRAS.386..887B}).  For example, \cite{2009arXiv0903.5315K}, recently found that quasar light curves have
characteristic time scales that are well-correlated with their black hole mass (estimated
from their emission line widths), but poorly correlated with their accretion rate (relative to
Eddington). A significant fraction of the quasars suitable for the analysis were the 
\cite{2003AJ....125....1G} variability-selected quasars in the LMC.

\begin{figure*}[p]
\centering
\includegraphics[width=8.5cm]{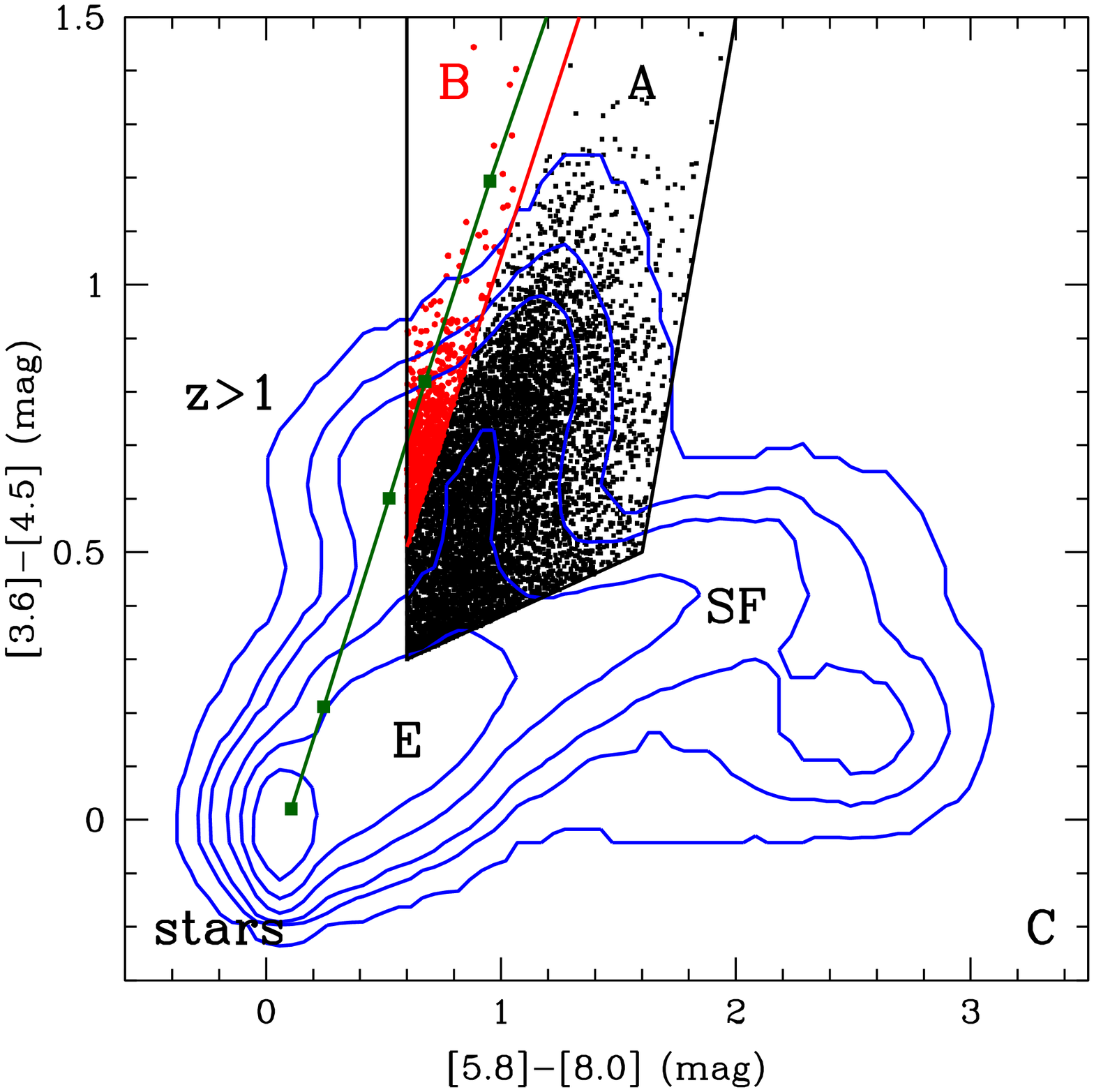}
\includegraphics[width=8.5cm]{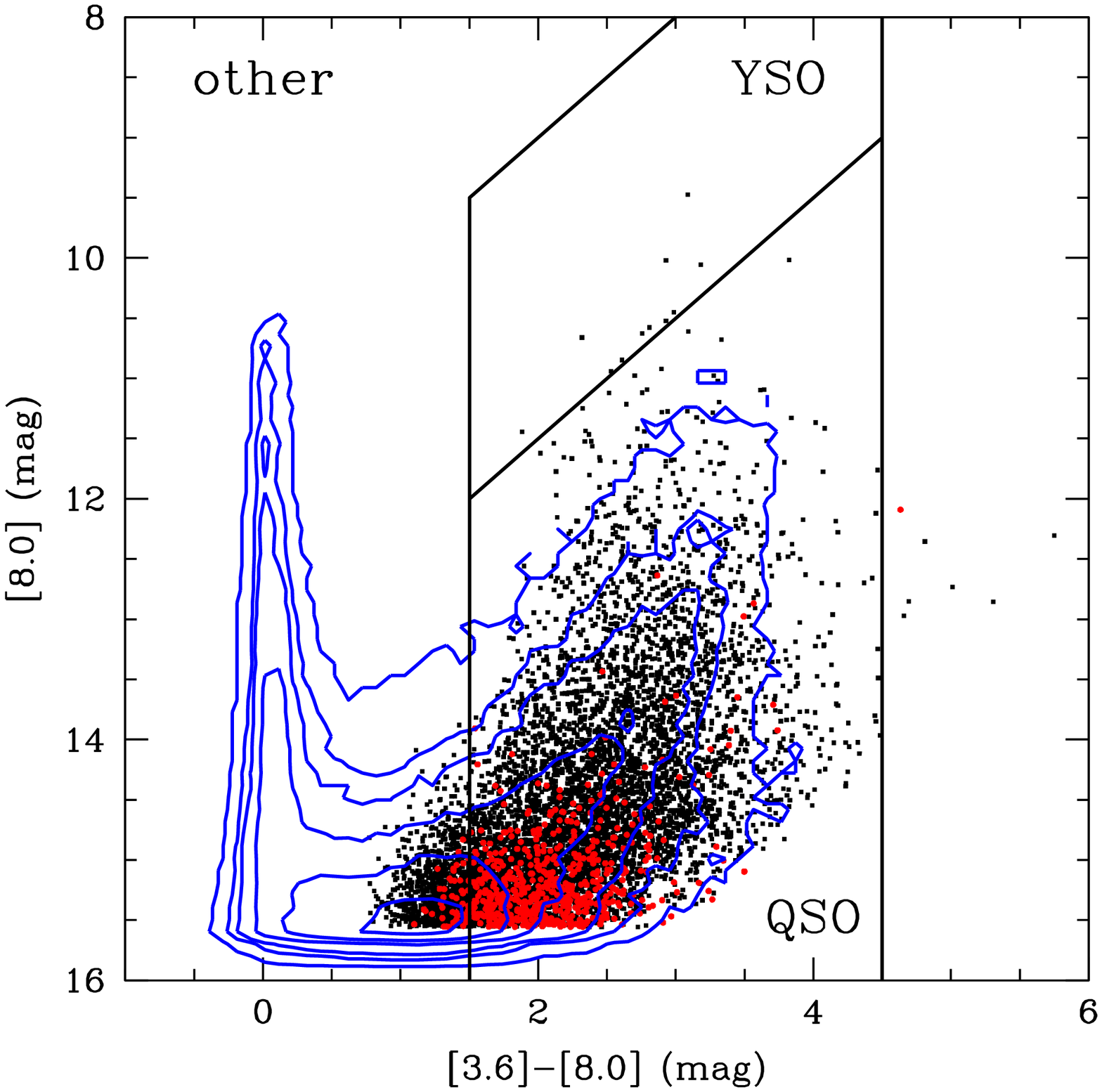}
\caption{$[3.6]-[4.5]$/$[5.8]-[8.0]$ color-color distribution (left) and $[3.6]-[8.0]$/$[8.0]$ CMD (right) of the extragalactic
  SDWFS. The smoothed contours are for 2, 10, 20, 50, 100 and 200 objects per bin with 0.1~mag bins for the colors
  and $0.2$~mag bins for the magnitude.  In the left panel we indicate the typical source type for the different color
  regions (stars, E -- ellipticals, SF -- star forming galaxies and $z>1$ galaxies).  We also show the Stern et al. (2005) AGN wedge and the color
  locus of black bodies (green curve marked with squares for temperatures of 1500, 1300, 1000, 800 and 600~K).
  In the right panel we show the boundaries of our YSO and quasi-stellar objects (QSO) selection regions (Cut \#2).
  The points are the AGN wedge objects, colored black if in region A and red if in region B
  (Cut \#1).  In the CMD, the region B AGN tend to be somewhat fainter than the region A AGN.  Few
  quasars lie in the YSO region, but this includes the brightest objects. }
\label{fig:bootes}

\centering
\includegraphics[width=8.5cm]{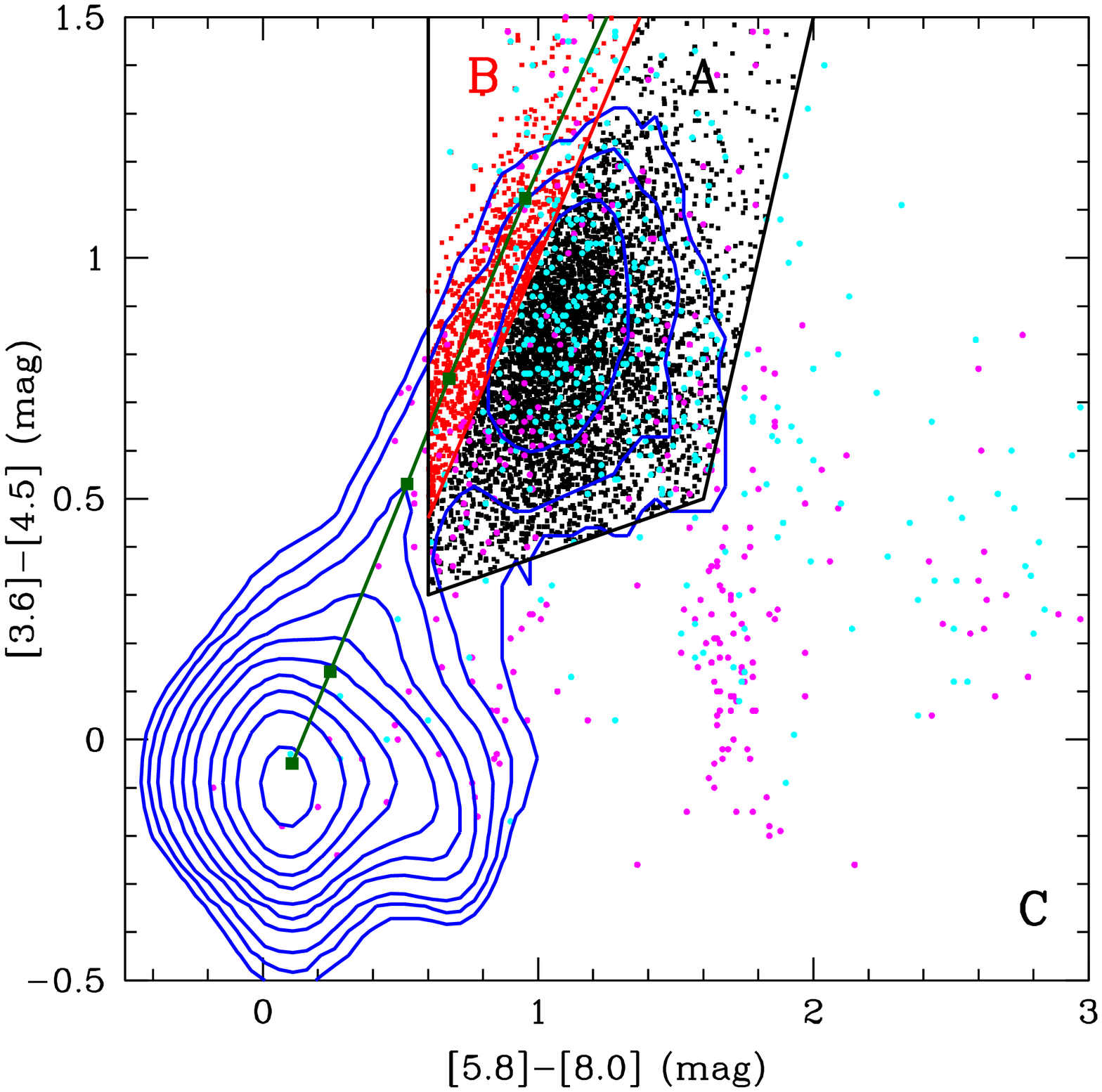}
\includegraphics[width=8.5cm]{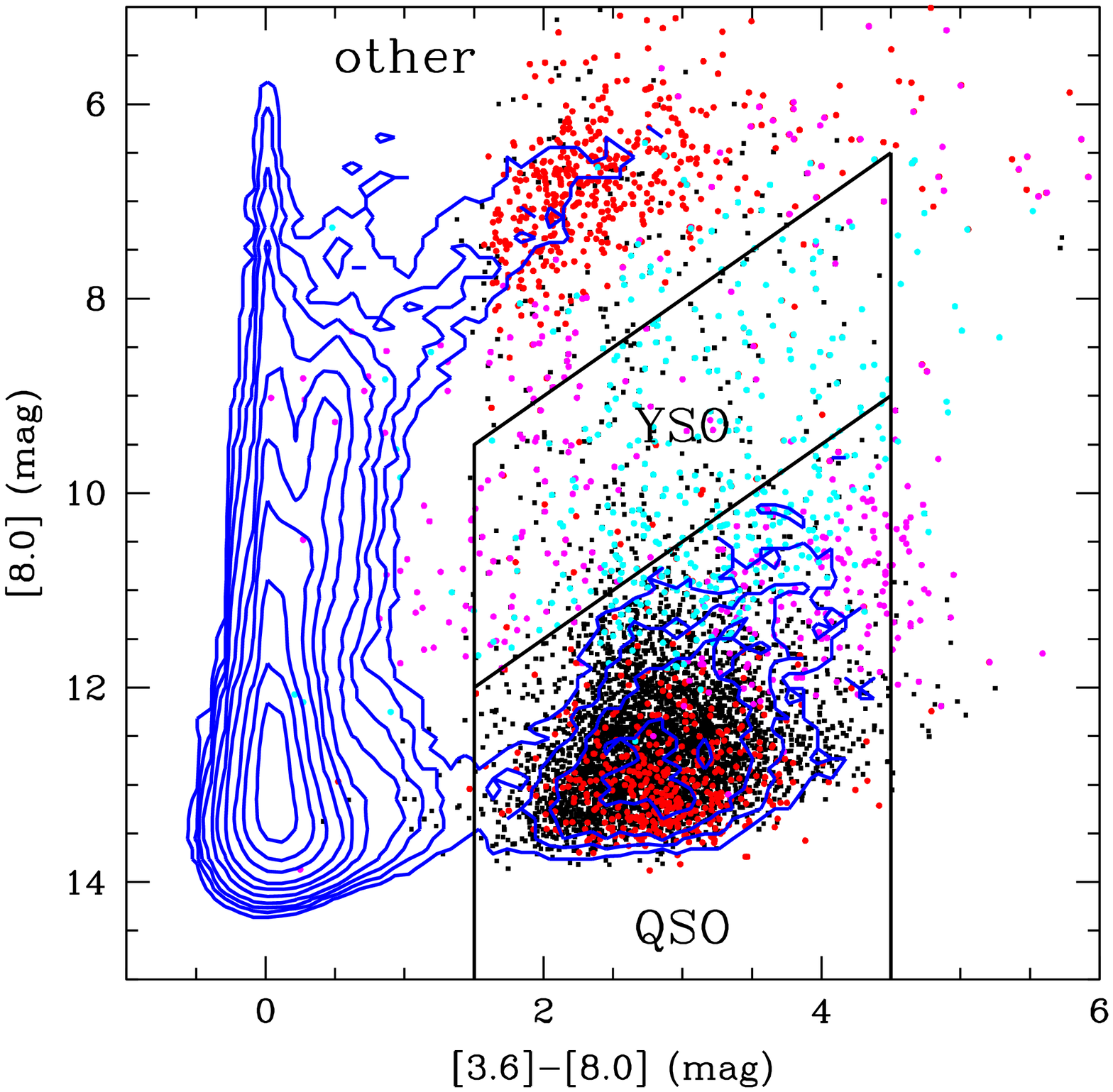}
\caption{The same distributions as in Figure~\ref{fig:bootes} but for the SAGE point source catalogs.  The 
smoothed contours are for 2, 5, 10, 20, 50, 100, 200, 500, 1000 and 2000 objects per bin where the bin
sizes are unchanged.  The black and red points again mark class A and B objects in the AGN wedge, but we
now see that class B objects are also found on the AGB sequence, but at apparent magnitudes very different
from those of AGN.  In both panels we also show confirmed (magenta) and highly probable (cyan) YSOs from 
\cite{2008AJ....136...18W}.  Many of these objects lie in the AGN wedge (478 of 722), and of those in
the AGN wedge 184 (191) are in the QSO (YSO) region.
}
\label{fig:sage}

\end{figure*}

Most quasars are identified by distinguishing their optical colors from those of stars,
with the SDSS samples representing the largest color-selected samples to date (e.g., \citealt{2009ApJS..180...67R}).  In dense stellar
fields, however, the enormously increased surface density of stars and stellar remnants make this
approach increasingly problematic.  Thus, searches in the Clouds and the Galactic Bulge,
mainly driven by the search for astrometric references, 
have focused on examining X-ray sources (\citealt{1997MNRAS.285..111T,2002ApJ...569..15D,2005A&A...442..495D}), 
radio sources (e.g., \citealt{1976MNRAS.174..259S, 1998A&AS..127..119F, 2002A&A...386...97J}), 
or time variability (\citealt{2002AcA....52..241E, 2003AJ....125....1G, 2005A&A...442..495D, 2005MNRAS.356..331S}). 
The largest samples come from the
time variability searches, but even samples of aperiodic objects with long term variability turn 
out to be dominated (80\%) by stars rather
than quasars (e.g., \citealt{2003AJ....125....1G}).  

Recent studies have shown that
mid-IR color selection is an extremely efficient means of finding luminous quasars
(\citealt{2005ApJ...631..163S,2004ApJS..154..166L}), although the method works poorly for low 
luminosity AGNs whose mid-IR emission is dominated by the host galaxy (see \citealt{2008ApJ...679.1040G}).
There are good reasons to expect this approach to work well even in crowded stellar fields with
significant visual extinction. Normal stars
have mid-IR (Vega) colors near zero, well away from the region of color space occupied by 
quasars---only the much rarer stars with significant emission from dust can begin to mimic the colors
of quasars and many of these will have optical to mid-IR colors or apparent magnitudes that are
inconsistent with those of quasars.  The principal, potential contaminants (see Blum et al. 2006),
are Asymptotic Giant Branch (AGB) stars,
planetary nebulae (PNe), and young stellar objects (YSO).

In Section~2 we combine results
from studies of the extragalactic NOAO Deep Wide Field Survey (\citealt{1999ASPC..191..111J}) B\"ootes field
based on the IRAC Shallow Survey (\citealt{2004ApJS..154..48E}), the {\it Spitzer} Deep Wide Field Survey (SDWFS, \citealt{SDWFS2009})
and the AGN and Galaxy Evolution Survey (AGES, C. S. Kochanek et al. 2009, in preparation) with the {\it Spitzer} SAGE Survey 
data (\citealt{2006AJ....132.2268M}) and OGLE-III optical survey 
(\citealt{2008AcA....58...69U,2008AcA....58...89U,2008AcA....58...329U}) of the LMC to develop a simple method of selecting AGN  
in these dense stellar fields.  We use this method to produce a catalog of 4699, and 657 quasar
candidates in the LMC and SMC based on the SAGE (\citealt{2006AJ....132.2268M}) and
S3MC (\citealt{2007ApJ...655..212B}) surveys. In practice, our basic result 
is very simple---the sources that \cite{2006AJ....132.2268M}, \cite{2006AJ....132.2034B}, \cite{2008AJ....136...18W} 
and \cite{2007ApJ...655..212B} systematically refer to as galaxies or star forming
galaxies are mostly relatively high ($z\gtorder 1$) redshift quasars.

%########################################################################
\section{The method}
\label{sec:data}

Our basic approach is explained by Figures~\ref{fig:bootes} and \ref{fig:sage} where we show mid-IR color-color and color-magnitude
distributions for both the SDWFS extragalactic field and the SAGE LMC field.  For the SDWFS
field we show all sources, while for the LMC we use the SAGE point source catalog containing $\sim4\times 10^6$ objects. 
In both cases we include only objects with magnitudes measured in all four IRAC bands
([3.6], [4.5], [5.8] and [8.0]), leaving us with 44,022 SDWFS and 220,158 SAGE sources.  
The SDWFS covers approximately 9~sq.~deg., while SAGE covers approximately 49 deg$^2$.
We start with the Stern et al. (2005) mid-IR color selection 
criterion (a.k.a. the AGN wedge) based on the $[5.8]-[8.0]$ and $[3.6]-[4.5]$ colors. 
The AGN wedge is known to be very efficient at separating luminous AGN from other extragalactic sources
and normal stars.

{\it Cut 1.} As we compare the color-color diagrams for the NDWFS and SAGE fields, we see that
for the LMC there is a plume of stars rising from the stellar peak (at Vega colors of zero) along the blue 
$[5.6]-[8.0]$ edge of the AGN wedge.  The plume lies along the color sequence for
500-1000~K black bodies.  For our first cut, we divide this color space into three
regions.  Class C consists of all objects outside the AGN wedge.  We divide the wedge
into Class A and B regions using the black body color sequence shifted by $0.2$~mag
in $[3.6]-[4.5]$ color.  The bluer class B objects are roughly consistent with
black bodies, while the redder class A objects are not.  In B\"ootes, 90\% of
AGN wedge sources are class A (6,599 versus 608).  If we now
examine where the class A and B objects lie in the $[3.6]-[8.0]$ versus $[8.0]$ 
color-magnitude diagram (CMD), we see that the class B objects lie either on the AGB sequence or are 
generally mixed with the class A objects.  Fortunately, there is a large flux
difference between the brightest quasars and Galactic or Magellanic Cloud AGB 
stars, which leads to our second cut.  

{\it Cut 2.}  Our second cut is based on the  $[3.6]-[8.0]$ versus $[8.0]$ CMD.  We
know from Cut 1 that we must include a criterion on the apparent magnitude to
eliminate AGB stars.  Moreover, if the dust emission from a star is 
characterized by a range of dust temperatures, then the star can lie in region A.
In particular, the SAGE (Whitney et al. 2008) criteria for candidate YSOs 
overlap region A but select apparent magnitudes brighter than the typical 
AGN.\footnote{When they examine their YSO candidates
more closely, Whitney et al. (2008) find they are dominated by a mixture of YSOs
and PNe. }
Thus, we divide the CMD into three regions.  We define the YSO region as
$1.5<[3.6]-[8.0]<4.5$ and $11.0 < [3.6] < 13.5$ based on Whitney et al. (2008)
and the fainter QSO region by $1.5 < [3.6]-[8.0] < 4.5$ and $[3.6] > 13.5$. The
remaining space is defined as ``other''.  From the B\"ootes field
we see that the vast majority of the quasars lie in the QSO region (6131 versus 11), 
but the bright quasars that are the best candidates for absorption line studies 
lie in the YSO region. Also, the YSO sample from Whitney et al. (2008) is biased toward brighter YSOs
and we expect some contamination from fainter YSOs in our QSO region.
While in B\"ootes we would lose 15\% of candidates due
to the $[3.6]-[8.0]$ color limit, this is not a factor in the brighter SAGE sample.  
Note that these cuts will work well for the Clouds and the Galaxy, with the balance
of the YSO region shifting in favor of quasars for Galactic fields, but they would need
to be significantly modified for a more distant field (e.g., M33) because the AGB
sequence would shift into the QSO region. 

\begin{figure}[t]
\includegraphics[width=8cm]{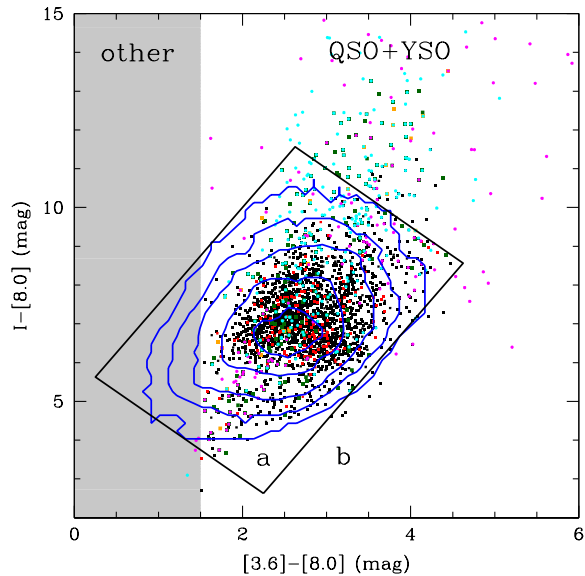}
\caption{$[3.6]-[8.0]$/$\hbox{I}-[8.0]$ color-color distributions.  The smoothed contours show the distribution of AGN   
  wedge objects in the SDWFS extragalactic survey (2, 5, 10, 20 and 30 objects per $0.1\times0.2$~mag color bin). The
  gray shaded region corresponds to the ``other'' region of Cut \#2.  The black, red, dark green and orange points
  show our QSO-A, QSO-B, YSO-A and YSO-B candidates from the SAGE survey, while the magenta and cyan points mark
  the confirmed and high probability YSOs from  \cite{2008AJ....136...18W}.  The box defines our Cut \#3 dividing
  the color space into regions a and b. 
  }
\label{fig:optical-IR}
\end{figure}

{\it Cut 3.} The OGLE-III survey covers roughly half of the SAGE survey area.  
We matched the SAGE sources with the OGLE-III optical photometry for
their region of overlap, and in Figure~\ref{fig:optical-IR} we examine the $\hbox{I}-[8.0]$/$[3.6]-[8.0]$ color-color
distribution of the B\"ootes quasars as compared to our QSO-A, QSO-B, YSO-A and YSO-B
sources as well as the ``high probability'' YSO candidates from Whitney et al. (2008).
We divide this color space into region ``a'' containing the B\"ootes quasars and 
region ``b'' outside of it.  Applying this cut excludes a reasonable fraction of YSOs
at the potential price of missing very high redshift ($z>6$) or obscured quasars.

A quasar candidate is any object other than class C objects from Cut 1 and ``other'' 
objects from Cut 2.  For the B\"ootes quasars, 6,599/608 are class A/B, 6,131/11 are
QSO/YSO, and 5,961/181 are class a/b.  Stellar contamination will be higher for
class B, YSO and b than for class A, QSO and a. Of the 722 Whitney et al. (2008)
YSO candidates, 478 are in the AGN wedge and, of these, 184 (191) lie in the QSO
(YSO) regions. Of those surviving the first two cuts in the QSO (YSO) regions,
147 (162) have I-band photometry and 123 (24) of these are in the ``a'' region of
Cut 3.

%########################################################################
\section{AGN behind the Magellanic Clouds }
\label{sec:results1}

When we apply the first selection cut to the SAGE catalog we find 5,402 objects in the AGN wedge 
(4,409 in region A and 993 in region B). When we apply the second cut, the QSO (YSO) region contains
3,891 (301) objects from region A and 464 (43) from region B for a total of 4,699 candidates.
Of these, we found OGLE-III optical counterparts 
for 1,981 (226) of the candidates in the QSO (YSO) region.  We label the candidates as QSO/YSO-[AB][0ab]
for their region in Cut 2, 1 and 3 respectively, where a 0 for Cut 3 means that we lack an
optical counterpart. There are 1,773 of the most promising QSO-Aa candidates (although, when matched to
an optical catalog, most of the $\sim$ 2,100 QSO-A0 candidates will become QSO-Aa candidates).
Table~\ref{tab:results_sage} presents our catalog of LMC quasar candidates.
We also applied the method to the much smaller $\sim$ 3 deg$^2$ area of the SMC--S3MC Survey
(\citealt{2007ApJ...655..212B}) again matching with the OGLE-III optical photometry.
Table~\ref{tab:results_s3mc} presents the catalog of SMC candidates, where we
found 526, 70, 45 and 16 QSO-A, QSO-B, YSO-A and YSO-B candidates and 508 with optical
matches.  We have not worried about extinction here, as it is generally too low to significantly
affect our mid-IR selection criterion, but statistical use of these samples will have to consider the effects
of extinction on the optical, spectroscopic limits, probably based on HI maps such as \cite{1998ApJ...503..674K}
or RR Lyrae (\citealt{arXiv:0905.3389}).

\begin{figure}[t]
\includegraphics[width=8cm]{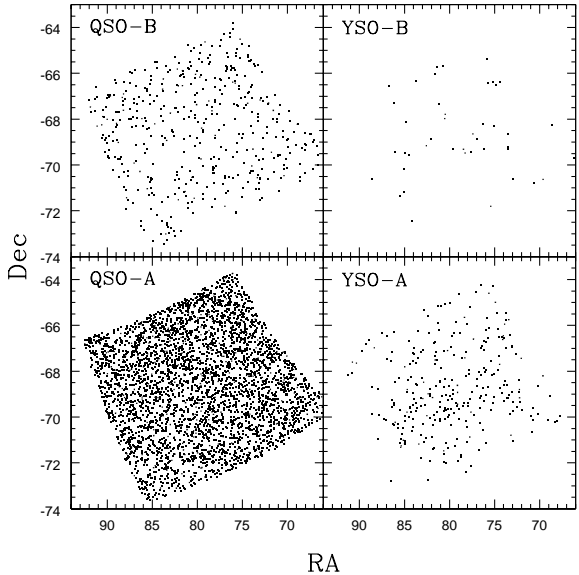}
\caption{The spatial distributions of the candidate classes. While the QSO-A sources are very uniformly
  distributed, there are signs of inhomogeneities and small scale clustering for the more contaminated
  YSO-A candidates. The QSO-B sources have a fairly uniform distribution and there are too few 
  YSO-B objects to draw any conclusions.
  }
\label{fig:spatial}
\end{figure}

The purity of the catalogs will depend on the selection regions.   Scaling the numbers of candidates in 
the B\"ootes field to the area and flux limits of the SAGE survey, we would expect 3,600, 16, 60 and 0 QSO-A, QSO-B, YSO-A 
and YSO-B candidates, respectively, compared to the 3,891, 464, 301 and 43 we identified as candidates
and the 155, 29, 164 and 27 YSO candidates in these regions from Whitney et al. (2008).  Taken at face
value, the QSO sample purities compared to an extragalactic field are 95\%, 3\%, 20\% and 0\% respectively
with the contamination in the QSO-A, YSO-A and YSO-B classes dominated by YSOs and that of the QSO-B
class dominated by an unidentified population.  Adding Cut \#3 leads to the loss of 2\%, 1\%, 19\% and
35\% of the candidates in these groups (50\%, 32\%, 64\% and 79\% of the \cite{2008AJ....136...18W}
YSO candidates), substantially increasing the expected purities.
Figure~\ref{fig:spatial} shows the spatial distribution of the sources.  The QSO-A 
population is fairly uniformly distributed across the SAGE region, while the more contaminated
QSO-B and YSO-A candidates show signs of inhomogeneities and clustering.  If the contamination
is dominated by YSOs, which tend to be clustered (see \citealt{2008AJ....136...18W}), then avoiding
candidates with close neighbors is likely to increase the purity of these samples.

We can test our efficiency on samples of known quasars in the region.  Of the 38 LMC and 9 SMC  MACHO variability quasars
(\citealt{2003AJ....125....1G}), 35 have complete mid-IR flux measurements and we classify 34 as
quasar candidates (97\%), missing one in the SMC quasars. 
\cite{2002ApJ...569..15D,2005A&A...442..495D}, largely based on OGLE data, 
identified 13 new quasars, 8 of which have mid-IR flux measurements, and we classify all 8 (100\%) as quasar 
candidates.  NED searches for all quasars near the LMC found 43 quasars with mid-IR fluxes 
and we classified 42 (98\%) as candidates.  The distribution of these quasars
(35, 2, 5 and 0 are QSO-A, QSO-B, YSO-A and YSO-B, respectively) by selection region is
very similar to the B\"ootes field.  The ROSAT PSPC source catalogs (\citealt{1999A&AS..139..277H,2000A&AS..142...41H})
have poor positional accuracies for crowded fields, but with our much smaller numbers of candidate quasars we
can attempt to find matches with some safety, as we clearly find that the X-ray sources are better correlated
with our candidates than with randomly selected SAGE/S3MC sources.  We note in the LMC (SMC) catalog the 117 (17) candidates within
the 90\% confidence position of a ROSAT source.  Based on comparisons to matches in random
catalogs of SAGE/S3MC sources, we estimate that these matches will have a false positive
rate of approximately 30\%.  

%########################################################################
\section{Summary and Conclusions}
\label{sec:conclusions}

We present a simple modification of the extragalactic mid-IR quasar selection
method of \cite{2005ApJ...631..163S} for use in dense stellar fields.  Only the
relatively rare, dusty stars are a source of contamination.  Of such
stars, AGB stars are relatively easily eliminated while sources like
YSOs and dusty PNe are significant contaminants when searching for the
rarer bright quasars or near the color locus of black bodies.
  We derive catalogs of 4,699 and 657 quasar candidates for the 
LMC and SMC using the SAGE (\citealt{2006AJ....132.2268M}) 
and S3MC (\citealt{2007ApJ...655..212B}) surveys, respectively.
We recover close to 100\% of the known quasars in the SAGE/S3MC survey
regions, demonstrating the efficiency of our selection method.  The purity of our sample
compared to a similar extragalactic search will be very high in the
primary QSO-A region, poorer if searching for the brightest
quasars in the YSO-A region, and very poor in the QSO/YSO-B regions
where the color locus of black bodies crosses the 
\cite{2005ApJ...631..163S} color selection region.  Fortunately, this
region of color space contains only 10\% of B\"ootes candidates.
We have not addressed contamination by other extragalactic sources, as
we know from our use of this approach in the B\"ootes field of 
the NDWFS (Stern et al. 2005, Gorjian et al. 2008)
that it is quite low.  In AGES, 69\% (84\%) of AGN wedge targets have $z>1$
($z>0.5$) and 86\% of the spectra are classed as broad-line AGN with only
1\% contamination by stars.

We examined the GLIMPSE survey (\citealt{2009PASP..121..213C}) of the
Galactic plane as well.  It is clear our approach will work, but the GLIMPSE
survey depth is markedly lower than those of the Clouds, so the numbers of
candidates are small.  Moreover, the survey is largely confined to regions 
of very high visual extinction where spectroscopic confirmation is infeasible.
Note, however, that the problem of stellar contamination diminishes in the Galactic 
plane relative to the Clouds because the dusty stars are shifted to lower
apparent magnitudes, moving them away from the quasars.  In lower stellar
density fields, like those of dwarf spheroidals; it would be straightforward
to use variants of our approach based on only the $[3.6]$ and $[4.5]$ bands
to efficiently identify background quasars with warm {\it Spitzer}.

Follow-up spectroscopy with AAOmega on the Anglo-Australian telescope 
(e.g., \citealt{2001MNRAS.328.1039C})
could efficiently confirm the candidates, since its large field of 
view and fiber number are well-matched to the source density. 
With an expansion in the number of extragalactic reference sources by
well over an order of magnitude, future improvements in the proper motions
of the Clouds will be limited by the time needed to make the proper motion
measurements rather than by the availability of reference sources.  Finding
quasars for absorption line studies will require working in the higher contamination YSO  
region of our selection method.

Perhaps most importantly, these are the most intensively monitored quasars
we presently have.  Many will have well-sampled light curves extending
for over a decade, allowing large statistical studies of the variability
properties of individual quasars rather than ensembles of quasars. 
Spectroscopy serves not only to determine the redshift but also to
supply an estimate of the quasar black hole masses based on their
emission line widths (e.g., \citealt{2000ApJ...533..631K}).  With large samples
it will quickly become clear if the tantalizing correlations between
variability properties and intrinsic properties noted by \cite{2009arXiv0903.5315K} 
hold generally.

%########################################################################
\acknowledgments

We thank Subo Dong, Kris Stanek, Jose L. Prieto and Roberto J. Assef 
for helpful discussions and comments on the project, and help with the SAGE and S3MC data.
We thank the anonymous referee for suggestions that improved the manuscript. 
This research has made use of the NASA/IPAC 
Extragalactic Database (NED) which is operated by the Jet Propulsion Laboratory, 
California Institute of Technology, under contract with the National Aeronautics 
and Space Administration.

\clearpage

\begin{landscape}
\LongTables

\begin{deluxetable}{lcccccccccccccccccc}
\tablecaption{AGN Candidates Behind the Large Magellanic Cloud\label{tab:results_sage}}
\scriptsize
\tablewidth{0pt}
\tablehead{
SAGE ID & RA & Dec & [3.6] &
$\sigma$[3.6] & [4.5] & $\sigma$[4.5] &
[5.8] & $\sigma$[5.8] & [8.0] & $\sigma$[8.0] & V & $\sigma$V & I & $\sigma$I & OGLE-III ID & type & z & Ref.}
\startdata
J042911.59-701649.5 & 67.29833 & $-$70.28043 & 15.54 & 0.06 & 14.82 & 0.11 & 13.98 & 0.08 & 13.01 & 0.06 & 99.99 & 9.99 & 20.77 & 0.27 & lmc158.6.1570 & QSO-Aa & -1 & -1 \\
J043014.94-701457.9 & 67.56225 & $-$70.24942 & 16.09 & 0.09 & 15.26 & 0.12 & 14.33 & 0.14 & 13.02 & 0.09 & 20.57 & 0.11 & 19.76 & 0.13 & lmc158.5.2597 & QSO-Aa & -1 & -1 \\
J043025.71-690419.1 & 67.60714 & $-$69.07199 & 15.98 & 0.05 & 15.35 & 0.08 & 14.48 & 0.11 & 13.60 & 0.08 & 21.40 & 0.16 & 20.31 & 0.21 & lmc156.5.1120 & QSO-Aa & -1 & -1 \\
J043036.29-690727.5 & 67.65124 & $-$69.12432 & 14.71 & 0.04 & 13.94 & 0.06 & 13.10 & 0.05 & 12.14 & 0.04 & 20.36 & 0.13 & 19.04 & 0.13 &  lmc156.6.398 & QSO-Aa & -1 & -1 \\
J043036.81-702131.2 & 67.65339 & $-$70.35866 & 15.51 & 0.07 & 14.69 & 0.07 & 13.73 & 0.10 & 12.72 & 0.05 & 20.99 & 0.21 & 19.79 & 0.16 & lmc158.6.2269 & QSO-Aa & -1 & -1 
\enddata
\tablecomments{The ``Type'' column gives the candidates' selection flags QSO/YSO-[AB][0ab] where $0$ for Cut 3 means that the source had no
  optical match because it was either outside the OGLE-III regions or too faint.  We also indicate if the source is a known quasar 
  (redshift and reference), if it is flagged as a YSO by Whitney et al. (2008),
  and if it is within the 90\% confidence radius of a ROSAT source (HP99, \citealt{1999A&AS..139..277H}).  We estimate that 30\% of the ROSAT 
  identifications will be false positives. For the ROSAT sources we indicate the ROSAT catalog number using the format HP99\_number.
  Unmeasured magnitudes and errors are indicated by 99.999 and 9.999 respectively. 
  One RA, Dec and magnitude digit has been truncated in this illustrative table in order to fit the page.
  }
\end{deluxetable}

\begin{deluxetable}{lcccccccccccccccccc}
\tablecaption{AGN Candidates Behind the Small Magellanic Cloud\label{tab:results_s3mc}}
\tablewidth{0pt}
\scriptsize
\tablehead{
S3MC ID & RA & Dec & [3.6] &
$\sigma$[3.6] & [4.5] & $\sigma$[4.5] &
[5.8] & $\sigma$[5.8] & [8.0] & $\sigma$[8.0] & V & $\sigma$V & I & $\sigma$I & OGLE-III ID & type & z & Ref.}
\startdata
J004255.06-732306.18 & 10.72944 & $-$73.38505 & 15.62 & 0.02 & 15.00 & 0.01 & 14.02 & 0.04 & 13.21 & 0.06 & 20.39 & 0.26 & 19.61 & 0.34 & smc125.2.30297 & QSO-Aa & -1 & -1 \\
J004255.51-732309.03 & 10.73132 & $-$73.38584 & 15.46 & 0.02 & 14.87 & 0.01 & 13.81 & 0.03 & 12.81 & 0.05 & 20.39 & 0.23 & 19.33 & 0.16 & smc125.2.30304 & QSO-Aa & -1 & -1 \\
J004258.33-732407.37 & 10.74308 & $-$73.40205 & 15.68 & 0.02 & 15.14 & 0.02 & 13.69 & 0.03 & 12.52 & 0.03 & 19.82 & 0.06 & 19.28 & 0.10 & smc125.2.30349 & QSO-Aa & -1 & -1 \\
J004323.89-732043.39 & 10.84956 & $-$73.34539 & 17.34 & 0.07 & 16.94 & 0.08 & 15.47 & 0.11 & 14.57 & 0.16 & 20.92 & 0.13 & 20.95 & 0.35 & smc125.2.41237 & QSO-Aa & -1 & -1 \\
J004413.65-724302.23 & 11.05690 & $-$72.71729 & 16.15 & 0.03 & 15.34 & 0.02 & 14.12 & 0.04 & 13.16 & 0.05 & 21.42 & 0.36 & 20.23 & 0.25 &  smc126.3.9008 & QSO-Aa & -1 & -1
\enddata
\tablecomments{We indicate if the source is a known quasar (redshift and reference), if it is flagged as a YSO by \cite{2007ApJ...655..212B},
and if it is within the 90\% confidence radius of a ROSAT source (H00, \citealt{2000A&AS..142...41H}).  For the ROSAT sources we indicate the
ROSAT catalog number using format H00\_number.
Unmeasured magnitudes and errors are indicated by 99.999 and 9.999 respectively.
  One RA, Dec and magnitude digit has been truncated in this illustrative
  table in order to fit the page.
}
\end{deluxetable}
\clearpage
\end{landscape}

\end{document}